# Superconducting phase diagrams of cuprates and pnictides as a key to the HTSC mechanism

K.V. Mitsen[1,] O.M. Ivanenko

*Lebedev Physical Institute, Russian Academy of Sciences, 119991, Moscow*

The article reviews experimental phase diagrams of cuprates and pnictides to demonstrate that specific features of the superconducting phase diagrams in both HTSC families can be understood within the framework of the proposed approach which assumes the formation, at heterovalent doping, of localized trion complexes consisting of a doped carrier and charge transfer (CT) excitons. The geometry of such cells containing CT excitons (CT plaquettes) in the basal plane of the crystal is determined by its crystal structure and type of dopant so that the dopant concentration range corresponding to the existence of a percolation cluster of CT plaquettes can be readily determined for each particular compound. These dopant concentration ranges coincide to good accuracy with the experimental ranges of superconducting domes on the phase diagrams of the HTSC compounds considered. The emergence of free carriers and the mechanism of superconducting pairing is related in this pattern to biexciton complexes (Heitler–London centres) emerging on neighbouring CT plaquettes.

## 1. Introduction

The nature of the normal state and the mechanism of superconductivity in two high-temperature superconductor families – cuprates and pnictides – still remains an issue of animated discussion. Undoped cuprates, as it follows from the experiment and as is supported by band calculations, possess electronic structures essentially other than pnictides. Still, despite the differences in their electronic structures, these compounds demonstrate a number of similar features. Among them, we would like to single out the following:

(1) *Low concentration of charge carriers*. Even at an optimal doping, the carrier concentration in cuprates and pnictides is lower than $10^{22}$ cm$^{-3}$, which corresponds to an average distance of $r_s > 0.4$ nm between the carriers and exceeds the distance between the anion and cation. This means that the interaction inside the cell is essentially unshielded, which enables the existence of well-defined charge transfer (CT) excitons [1].

(2) *High ionicity of cuprates and pnictides*, which suggests a large contribution of Madelung volume energy $E_M$ to the electronic structure of the basal planes and the possibility to locally change the electronic structure by doping with localized carriers.

(3) *Similarity of cuprates' and pnictides' phase diagrams*. As is known, the predominant majority of undoped cuprates and pnictides are not superconductors. Superconductivity in them emerges as the result of heterovalent doping, i.e., at a partial substitution of an atom for another atom with higher or lower valency. Herewith, the superconductivity region on the phase diagrams of these two classes is restricted by a certain concentration interval $(x_1;x_2)$, i.e., only within this doping interval there are conditions for the superconductivity mechanism to be activated. The dependence of $T_c(x)$ within the interval $(x_1;x_2)$ is, in the general case, dome-shaped (Fig. 1) with a maximum at $x_{opt}$.

---
[1] E-mail: mitsen@sci.lebedev.ru

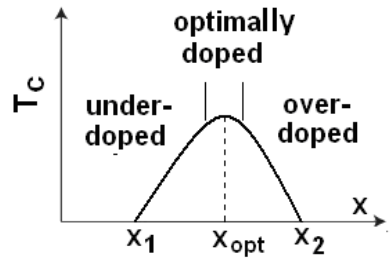

Figure 1. (a) A typical superconducting phase diagram of doped cuprates and pnictides. Superconductivity takes place at a dopant concentration of $x$ within the range of $(x_1;x_2)$ with a $T_c$ maximum at $x = x_{opt}$.

The region of concentrations in the vicinity of $x_{opt}$ corresponding to the region with maximal $T_c$ is conventionally taken to be the region of optimal doping, whereas the concentration regions on the left- and right-hand sides of the optimal doping region are named, respectively, the underdoping and overdoping regions.

Figure 2 shows experimental superconducting phase diagrams of the most well investigated HTSCs, cuprates and pnictides. As seen in the figure, concentrations $x_1$, $x_{opt}$ and $x_2$ essentially differ for different HTSCs. Still, evident similarity of the superconducting phase diagrams for cuprates and pnictides enables an assumption of some common and sufficiently rough mechanism that acts irrespective of the fine details of the band structure and provides for the superconducting pairing in these materials.

The aim of this review is to show that the values of $x_1$, $x_{opt}$ and $x_2$ for various HTSCs can be calculated based on the superconducting pairing mechanism common for cuprates and pnictides. Herewith, in a number of cases, when superconducting phase diagrams demonstrate fine features (narrow dips, jumps etc.), the positions of these features can be reproduced with unprecedented accuracy.

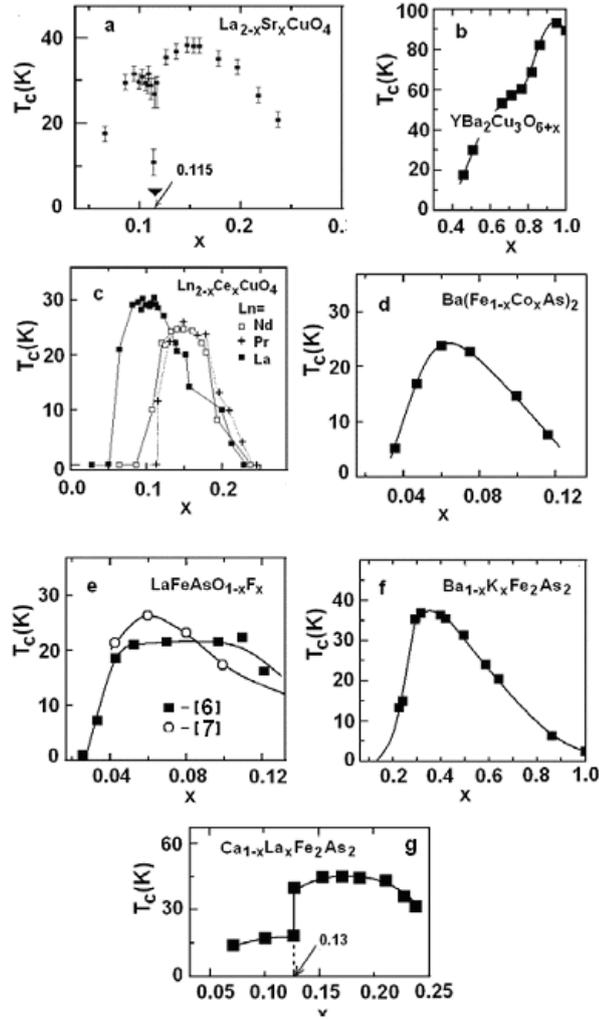

Figure 2. Experimental phase diagrams of some cuprates and pnictides:
(a) $La_{2-x}Sr_xCuO_4$ [2]; (b) $YBa_2Cu_3O_{6+x}$ [3]; (c) $Ln_{2-x}Ce_xCuO_4$ [4]; (d) $Ba(Fe_{1-x}Co_xAs)_2$ [5]; (e) $LaFeAsO_{1-x}F_x$ [6,7]; (f) $Ba_{1-x}K_xFe_2As_2$ [8]; (g) $Ca_{1-x}La_xFe_2As_2$ [9].

## 2. Mechanism of HTSC heterovalent doping

Let us consider the specific features of HTSC materials under heterovalent doping. The heterovalent-doping technique is broadly used in semiconductor studies to establish a given concentration of certain-sign carriers in the bulk of the crystal. In semiconductors, the concentration of dopants $N_d \sim 10^{18}$–$10^{19}$ cm$^{-3}$ is in most cases sufficient to achieve a degenerate distribution of carriers at which their concentration and, correspondingly, Hall constant, do not depend on temperature. At the same time, cuprates and pnictides, even at optimal doping when the dopant concentration is $10^{20}$–$10^{21}$ cm$^{-3}$, are observed to have a strong dependence of the Hall constant on temperature [10–12] and no direct proportionality of the Hall concentration of carriers and dopant concentration. What is more, quite a number of experiments using various techniques obtained convincing proofs of the localization of doped carriers in the nearest vicinity of the dopants [13–17]. That is, the experiment demonstrates that the doped carriers' localization regions do not overlap.

These facts, if recognized, pose several questions:

(1) What is the mechanism of doped carriers' localization?
(2) Where does superconductivity come from and what does the HTSC superconducting phase diagram reflect in this case?

(3) What is the mechanism of free carriers' generation?
(4) What is the mechanism of superconductivity?

As an answer to the first question, we consider a mechanism of HTSC heterovalent doping, in which doped carriers are self-localized as the result of the formation of trion complexes. In a this complex, a doped carrier is bound by Coulomb interaction with charge-transfer (CT) excitons that are generated in the vicinity of the carrier due to a local deformation of the electronic structure. The proposed mechanism, as we will show, can lead to (in the general case) the formation of an inhomogeneous phase whose properties coincide with the experimentally observed properties of HTSCs.

Figures 3a,b show an arrangement of anions and cations (their projections, to be more exact) in the basal planes of cuprates and pnictides. In cuprates, Cu and O ions are in the same plane; in pnictides, Fe ions are in the plane and As ions are at the vertices of the regular tetrahedra such that their projections form a square sublattice in the basal plane.

Cuprates in undoped states are antiferromagnetic Mott insulators in which the empty subband of copper $3d^{10}$ states is separated from the filled oxygen O2p band by a gap $\Delta \sim 2$ eV. In undoped pnictides that are antiferromagnetic (bad) metals, states on the Fermi surface are formed mainly by Fe orbitals, whereas electron states on As are ~2 eV lower than $E_F$ [18–20]. Thus, *to transfer an electron from an anion (O, As) to a cation (Cu, Fe), approximately the same energy of ~2 eV should be spent both in cuprates and pnictides.*

Figures 3c,d show schematically the band structures of (c) undoped cuprates (in electron representation) and (d) undoped pnictides (in hole representation). As it follows from the figures, energy $\Delta_{ib}$ required for interband transition related to the transfer of an electron from an oxygen ion to a copper ion (in cuprates) (Fig. 3c) or the transfer of a hole from an iron ion to arsenium ion (in pnictides) (Fig. 3d) is in both cases approximately the same, $\Delta_{ib} \sim 2$ eV.

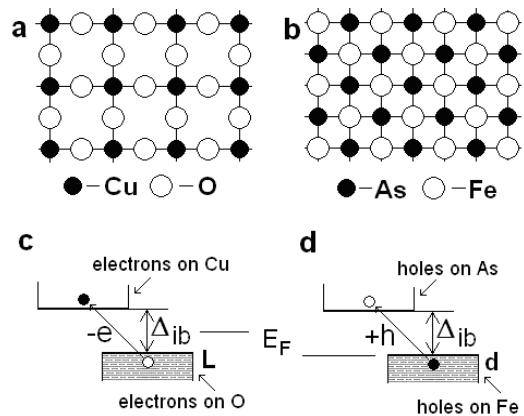

Figure 3. (a,b) Arrangement of the projections of anions and cations in the basal planes of cuprates (a) and pnictides (b); (c) the band structure of undoped cuprates in electron representation; (d) the band structure of undoped pnictides in hole representation. Energy $\Delta_{ib}$ required for interband transition related to the transfer of an electron from an oxygen ion to a copper ion (in cuprates), or to the transfer of a hole from an iron ion to an arsenium ion (in pnictides), is approximately the same in both cases and makes ~2 eV.

At the same time exciton-like excitation is also possible, which has a lower energy $\Delta_{ct} < \Delta_{ib}$ and corresponds to the local transfer of an electron (hole) from an anion (cation) to the nearest cation (anion) (Fig. 4a,b) to form the bound state (of a CT exciton).

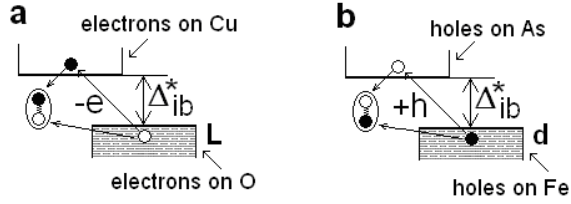

Figure 4. Formation of a CT exciton in cuprates (a) and pnictides (b). To form the CT exciton, the band gap $\Delta_{ib}$ should be reduced to $\Delta_{ib} = \Delta^*_{ib} < E_{ex}$, where $E_{ex}$ is CT exciton binding energy.

If $\Delta_{ib}$ is somehow gradually decreased, we shall first arrive to a state with $\Delta_{ib} = \Delta^*_{ib}$, in which $\Delta_{ct} = 0$ (i.e., $D^*_{ib} \leq E_{ex}$, where $E_{ex}$ is CT exciton binding energy). If $\Delta_{ib}$ is decreased homogeneously, the condition $\Delta_{ct} = 0$ will be satisfied for the entire basal plane. If we locally suppress $\Delta_{ib}$ in some regions, a continuous cluster of the phase with $\Delta_{ct} = 0$ will emerge at an excess of the percolation threshold over those regions.

Let us have a phase with $\Delta_{ct} = 0$ (i.e., $\Delta^*_{ib} \leq E_{ex}$). This means that two-particle transitions to and fro become possible between two one-particle states (L electron + d hole) on the one hand and an exciton two-particle state (d electron + L hole) on the other hand. For this reason, in a phase with $\Delta_{ct} = 0$ electron (hole) states in L (d) bands should be considered as a superposition of band and exciton states.

The events corresponding to the generation of a CT exciton (localized in a unit cell) are, in cuprates, the occurrence of an electron on the central Cu cation and a hole distributed over four surrounding anions of O; in pnictides, a hole on the As anion and an electron distributed over four surrounding cations of Fe. This hydrogen-like ionic complex, for which the condition $\Delta_{ct} = 0$ is satisfied and in which a CT exciton, resonantly interacting with the band states, can be formed, will be called a CT plaquette. In this complex, the state of an electron (in cuprates) and a hole (in pnictides) should be considered as a superposition of band and exciton states.

Let us consider now how the doping transforms the band structure of parent phases of both cuprates and pnictides (Fig. 5), thus leading to the formation of CT plaquettes.

As the gap $\Delta_{ib}$ in cuprates and pnictides is largely determined by Madelung energy $E_M$, we need to locally decrease the value of $E_M$ to form a CT plaquette centred on a given (Cu or As) ion. Hypothetically, this can be done by arranging additional localized charges of respective value and sign either on the central (Cu or As) ion or on one of four surrounding ions (of O or Fe), or else in the immediate vicinity from them. Interestingly, exactly this mechanism of decreasing $\Delta_{ib}$ appears to be realized in HTSCs under doping.

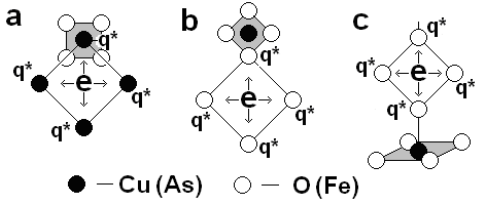

Figure 5. Formation of CT plaquettes and self-localization of doped charges in cuprates and pnictides. The CT plaquette is shaded.
(a) Hole doping in pnictides or electron doping in cuprates.
(b) Hole doping in cuprates or electron doping in pnictides.
(c) A dopant and a doped charge outside the basal plane (YBCO, BSCO etc.).

Introduction of a dopant ion into the lattice is accompanied with the emergence (in the vicinity of the dopant projection) of an additional carrier (an electron or a hole) in the basal plane or outside it. This doped carrier, in accordance with the neighbourhood symmetry, imparts fractional charge $q^* \approx \pm|e|/4$ to corresponding orbitals of each of the nearest ions. At the hole doping, holes are distributed over four nearest ions of O or As, and electrons, at the electron doping, over four nearest ions of Cu or Fe.

Charge $q^*$ emerging on each of four Cu (As) ions (Fig. 5a) locally decreases Madelung energy. Herewith, we neglect the effect of the charge of the dopant, which is at a larger distance. In turn, a local decrease of Madelung energy reduces the gap $\Delta_{ib}$ for the transition of an electron (hole) to each of four ions of Cu (As) from the surrounding ions of O (Fe). As estimates show [21], at $q^* \approx |e|/4$ and the existing interatomic distances this decrease is ~1 eV and, in our assumption, is sufficient to fulfil the condition $\Delta_{ct} = 0$, i.e., to form CT plaquettes on these ionic complexes.

As each CT plaquette is bound by Coulomb attraction with doped charge $q^*$, the doped carrier self-localizes in the nearest vicinity of the dopant, which is supported by the results of [13–17].

Thus, the localization boundary of the doped carrier is determined by the condition that the same charge $q^*$ (~$\pm|e|/4$) sufficient for CT plaquette formation be at this boundary per each anion (cation). In fact, we produce a trion complex where the doped carrier is bound to CT excitons that resonantly interact with the band states. This mechanism of CT plaquette formation (Fig. 5a) is realized in electron-doped cuprates and hole-doped pnictides.

In the case shown in Fig. 5b, doped charge $q^*$ occurring at each of four O (Fe) ions locally decreases Madelung energy and forms a CT plaquette in the next ion square centred on the nearest Cu (As) ion, because only these ions can be CT plaquette centres. This scheme is realized for hole-doped cuprates and electron-doped pnictides.

Figures 5a,b represent the cases when a doped carrier is in the basal plane. However, it is possible that doped charge $q^*$ localizes outside the plane (Fig. 5c), e.g., on apical ions of oxygen, as it takes place in YBCO and other two-plane cuprates. This charge also decreases electron energy on the nearest Cu ion by the required value and thereby forms a CT plaquette in the plane of $CuO_2$.

If, in the cases shown in Figs 5a,b, the dopant concentration is increased until the neighbouring trion complexes begin to overlap each other, this will lead to the delocalization of doped carriers and, correspondingly, to the transition to the overdoped phase. In the case of Fig. 5c, transition to the overdoped phase with the doping level increasing will occur when the calculated charge $q^*$ per apical cation will significantly exceed the required value of $|q^*| \sim |e|/4$. The latter takes place, in particular, in the HTSC $Ba_2Sr_2CaCu_2O_{8+x}$ ($0 < x < 2$), where higher doping levels can be achieved as compared with $YBa_2Cu_3O_{6+x}$ ($0 < x < 1$).

According to the above said, CT plaquettes limiting the propagation of the doped carrier are formed around the dopant projection in the basal plane. In the case shown in Fig. 5c, when the doped carrier is localized outside the basal plane, two CT plaquettes are formed, one in each plane; in the cases shown in Figs. 5a,b, the number of CT plaquettes formed around the doped carrier localization region depends on the neighbourhood symmetry and, as it will be seen further on, can be from 4 up to 8.

Thus isolated localization regions of doped carriers enclosed by CT plaquettes are formed in the basal plane of doped HTSCs at sufficiently low dopant concentrations. If we accept this pattern, a question arises: what do the superconducting phase diagram of an HTSC and its dome-like shape reflect? If $0 < x_{opt} < 1$, i.e., the superconducting phase exists in an intermediate concentration range, then the only possibility to explain the position of the dome and its shape within the framework of the model is to relate the superconducting phase with the CT plaquette phase. That is, to assume that the superconductivity takes place in the percolation cluster that unites CT plaquettes. A plaquette cluster should be considered as a continuous network of adjacent cations (in cuprates) or anions (in pnictides) that are the centres of CT plaquettes. A CT-phase cluster can consist of two or more CT plaquettes (see Sections 3 and 4 below). As a variant of such a cluster, we will also consider a plaquette network where small plaquette clusters that comprise two or more CT plaquettes are linked by regions of doped carriers' localization.

Further, we will determine the existence regions of percolation clusters from CT plaquettes in the basal planes of various HTSC compounds and will compare them with the concentration ranges of dopants corresponding to the existence of high-temperature superconductivity on the phase diagrams of those compounds.

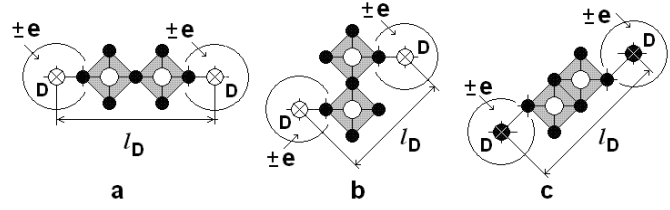

Figure 6. Examples of the arrangement of dopant (D) projections in the basal plane that lead to the formation of a chain of CT plaquettes (shaded). The circles around the dopant projections conventionally designate the boundaries of the doped charges' localization regions.

In the cases when a doped charge is in the basal plane, for the percolation network of CT plaquettes to be formed, it is necessary that the dopant projections be at some fixed distances $l_D$ from each other forming clusters of CT plaquettes. The percolation network in the basal plane can be formed if the dopant projections are ordered into a square superlattice with parameter $l_D$. For this reason, a large role in the formation of the superconducting phase is played by dopants' ordering that provides for the possibility of forming percolation clusters of CT plaquettes. We believe that, in doped cuprates and pnictides, there is a 3D ordering of localized carriers (projections of dopants) into 3D lattices with parameter $l$ determined by the dopant concentration (with site occupation $\nu \leq 1$). This ordering, in our opinion, is due to the emergence of elastic deformations at the difference of sizes of dopant and matrix atoms and/or due to the orientational interaction of dipoles (dopant ion + localized doped carrier) [22]. In the formed superlattice, the dopant projections are assumed to be randomly distributed.

Let us have a domain in which the dopant projections are randomly distributed over the nodes of a superlattice with parameter $l_D$. The dopant concentration $x_{max} = 1/l_D^2$, corresponding to the full filling of lattice nodes, will be taken for the upper boundary of the optimal doping region. This boundary corresponding to the maximal concentration of CT plaquettes also corresponds, according to our assumption, to the maximum $T_c$. For the lower boundary, $x_{min} = 0.593/l_D^2 \approx 0.6/l_D^2$, we will conventionally take a dopant concentration corresponding to the site percolation threshold on a square lattice with parameter $l_D$ [23]. This choice is determined by the fact that the existence of physically significant domains with the percolation network of dopant projections, spaced by a distance of $l_D$ from each other, is possible only at

$$0.6/l_D^2 < x < 1/l_D^2.$$

This does not mean that the percolation cluster on a lattice with parameter $l_D$ should occupy the entire crystal, but only that such domains will exist within only this concentration range. Herewith, a superconductivity in the entire crystal can emerge owing to the Josephson coupling between such domains. A particular shape of the curve of $T_c(x)$ is determined by the possibility of realizing various types of dopant ordering in this crystal structure [24,25].

Now we are passing on to the consideration of particular compounds.

## 2.1. $La_{2-x}Sr_xCuO_4$

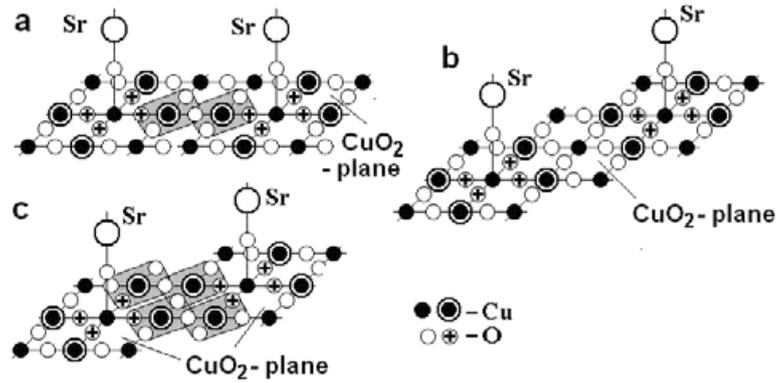

Figure 7. Formation of the percolation network of CT plaquettes (shaded) in $La_{2-x}Sr_xCuO_4$ at various distances between dopant projections: (a) $l_D = 3a$; (b) $l_D = a\sqrt{8}$; (c) $l_D = a\sqrt{5}$. Double circles are Cu ions, which are the centres of CT plaquettes.

In the case of $La_{2-x}Sr_xCuO_4$, the doped hole emerging at the substitution of $Sr^{2+}$ for $La^{3+}$ is in the $CuO_2$ plane (Fig. 7) and is distributed over four oxygen ions pertaining to the oxygen octahedron adjacent to Sr ion [13,14]. Each of four fractional charges $q^*$ on oxygen ions forms CT plaquette in the next ion square centred on the nearest Cu cation (Fig. 5b). It is readily seen that only two variants of the relative arrangement of two nearest projections of Sr on the $CuO_2$ plane are possible so that they could form a pair of neighbouring CT plaquettes (Figs. 7a and 7c). These cases correspond to two possible distances between them, $l_D = 3a$ and $l_D = a\sqrt{5}$ ($a$, lattice parameter). The pairs of CT plaquettes formed in the $CuO_2$ plane centred on Cu ions (double circles) are encircled with solid lines. Note that in an intermediate case when the distance between the Sr projections is $l_D = a\sqrt{8}$ (Fig. 7b), no pairs of CT plaquettes are formed on neighbouring Cu ions. In accordance with two different variants of dopant projections' arrangement, the phase diagram of $La_{2-x}Sr_xCuO_4$ is expected to have two regions of optimal doping at $0.066 < x < 0.11$ and at $0.12 < x < 0.2$ (corresponding to the ordered arrangement of Sr projections onto 3x3 and $\sqrt{5} \times \sqrt{5}$ lattices). Note that the experimental value of the upper optimal concentration, $x = 0.15$ (optimal in the sense of the magnitude of $T_c$) differs from the expected value, $x = 1/5$, though a jumplike decrease of the volume of the superconducting phase is observed namely at $x = 1/5$ [26]. This discrepancy can be explained by the formation of clusters of normal metal, which begins at $x > 0.15$ [27]. Within the interval of $0.11 < x < 0.12$, domains with the percolation network of CT plaquettes cannot exist. This is the so-called 1/8 anomaly, which, however, takes place not at $x = 0.125$ but, according to this consideration, at $x = 0.115$, in total agreement with experimental data (Fig. 2b) [2,28].

## 2.2. $Ln_{2-x}Ce_xCuO_4$

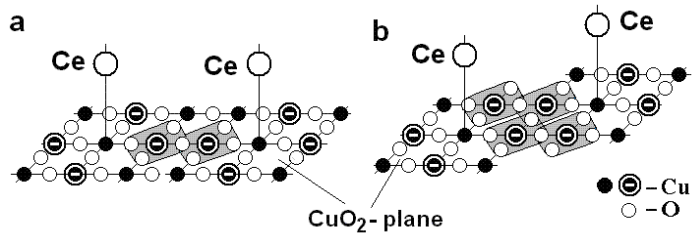

Figure 8. Formation of CT plaquettes (shaded) in $Ln_{2-x}Ce_xCuO_4$ at various distances between dopant projections: (a) $l_D = 3a$; (b) $l_D = a\sqrt{5}$; double circles are Cu ions, which are the centres of CT plaquettes. Clusters of CT plaquettes are shaded.

In the electron-doped HTSC $Ln_{2-x}Ce_xCuO_4$ the doped electron emerging at the substitution of $Ce^{4+}$ for $Ln^{3+}$ is in the $CuO_2$ plane (Fig. 8) and is distributed over four Cu ions. In accordance with Fig. 5a, four external ions of Cu are the centres of CT plaquettes. Similar to $La_{2-x}Sr_xCuO_4$, two types of dopant projections' arrangement and, respectively, two regions of optimal doping – $0.066 < x < 0.11$ and $0.12 < x < 0.2$ – where a network of CT plaquettes can be formed, are possible in this compound. However, on the phase diagrams of $Nd_{2-x}Ce_xCuO_4$ and $Pr_{2-x}Ce_xCuO_4$ only one superconducting dome is observed with a $T_c$ maximal at $x = 0.15$ and falling to zero at an increase of $x \to 0.2$. We associate this discrepancy with a low degree of order of Ce ions in the lattice owing to the proximity of the atomic radii of Nd, Pr and Ce ($r_{Nd} \approx r_{Pr} \approx r_{Ce} \approx 0.185$ nm [29]). At the same time, on the phase diagram of $La_{2-x}Ce_xCuO_4$ ($r_{La} \approx 0.195$ nm [29]) one can clearly see two superconducting domes in the expected intervals with the local $T_c$ maxima at $x = 0.11$ and $x = 0.15$, as in $La_{2-x}Sr_xCuO_4$ (Fig. 2c [4]).

## 2.3. $YBa_2Cu_3O_{6+x}$

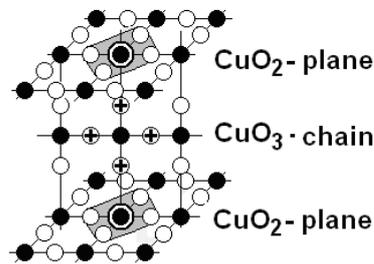

Figure 9. Formation of CT plaquettes in $YBa_2Cu_3O_{6+x}$. Provided that two positions in the chains in succession are filled with oxygen ions, one hole, distributed over four ions of the oxygen square, emerges in this square. Herewith, charge $\approx q^*$ ($\approx +|e|/4$) is imparted to the apical oxygen ions nearest to the Cu ion. This is sufficient to form a CT plaquette in each $CuO_2$ plane. (CT plaquettes are shaded.)

The parent compound $YBa_2Cu_3O_6$ is doped by introducing excess oxygen $\delta$ into the plane of the chains. In the case when two positions in the chain in succession are occupied by oxygen ions (Fig. 9), an oxygen square forms, with one hole distributed over four oxygen ions of this square (circled plus symbols). Herewith, additional positive charges $\approx q^*$ ($\approx +|e|/4$) emerge on apical oxygen ions nearest to the in-plane Cu ions (double circles), which results in the formation of CT plaquettes with the centre on these Cu ions (Fig. 5c). As the concentration of oxygen pairs in a chain is equal to $x^2$, the percolation cluster of the CT plaquette phase will exist at $x^2 > 0.593$, i.e., at $x > 0.77$. Thus, the region of optimal doping for $YBa_2Cu_3O_{6+\delta}$ is within the interval of $0.77 < \delta < 1$, in accordance with the experiment (Fig. 2a [3]). We should note that YBCO is the only case, known to us, when doping can realize condition $\Delta_{ct} = 0$ for the entire basal plane. In the case of, e.g., another double-plane cuprate $Bi_2Sr_2CaCu_2O_{8+x}$, where $0 < x < 2$, inhomogeneous filling of excess oxygen positions leads to various charges $q^*$ on the apical ions of oxygen, a consequence of which is the co-existence of underdoped and overdoped regions in one basal plane along with optimally doped regions.

## 2.4. Ba(Fe$_{1-x}$Co$_x$As)$_2$

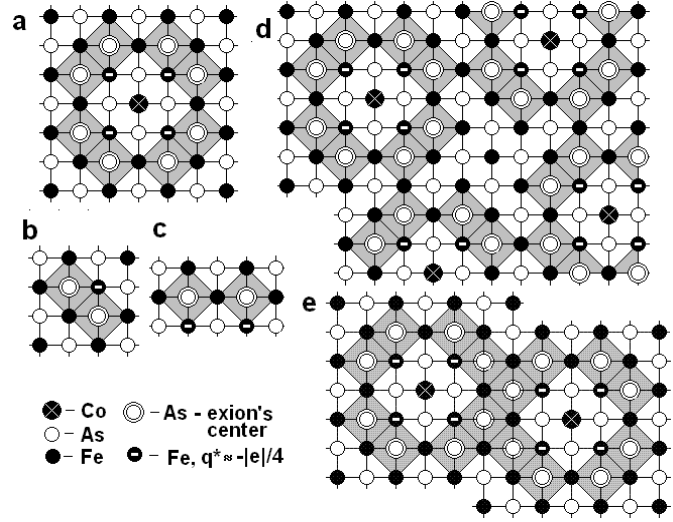

Figure 10. Formation of a cluster of CT plaquettes in Ba(Fe$_{1-x}$Co$_x$As)$_2$. Open circles indicate projections of As ions at the vertices of the tetrahedra onto the Fe plane; (a) formation of CT plaquettes (shaded) around the Co ion; (b,c) possible variants of combining CT plaquettes into a continuous cluster; the variants differ by different arrangements of Co ions' projections at the same distance between As anions; (d,e) formation of a percolation cluster of CT plaquettes on $\sqrt{20} \times \sqrt{20}$ and $\sqrt{13} \times \sqrt{13}$ square lattices.

In this compound, doping is performed by substituting Co atoms for Fe in the basal plane (Fig. 10). An additional electron emerging at the substitution imparts charge $q^*$ ($\approx -|e|/4$) to each of four surrounding Fe ions (Fig. 10a). In principle, two variants of combining CT plaquettes into a continuous cluster (Figs. 10b,c) are possible that differ by different arrangements of Co ions' projections at the same distance between As anions. At an ordered arrangement of dopants in a $\sqrt{20} \times \sqrt{20}$ lattice we will have a percolation network of CT plaquettes (Fig. 10d). A corresponding concentration of optimal doping is $x = 0.05$, and the concentration corresponding to the percolation threshold is $x = 0.03$, which is in good agreement with the experimental phase diagram of Ba(Fe$_{1-x}$Co$_x$As)$_2$ (Fig. 2d). It is readily seen that other distances between dopants, ensuring the formation of a percolation cluster of CT plaquettes, are possible within the interval of $a\sqrt{13} \leq l_D \leq a\sqrt{20}$, too (Fig. 10e). This interval corresponds to the concentration range of $0.03 \leq x \leq 0.08$ (Fig. 2d [5]).

## 2.5. LaFeAsO$_{1-x}$F$_x$

In this compound, at the substitution of fluorine for oxygen, one electron is doped into the basal plane. The projection of F ion onto the basal plane coincides with the position of Co ion in Fig. 10a. As a consequence, the symmetry of doped charge distribution in LaFeAsO$_{1-x}$F$_x$ will be similar to that in Ba(Fe$_{1-x}$Co$_x$As)$_2$ (Fig. 10a). Therefore, the phase diagrams of LaFeAsO$_{1-x}$F$_x$ and Ba(Fe$_{1-x}$Co$_x$As)$_2$ in respective concentration ranges should coincide, which is consistent with the experiment (Fig. 2e [6,7]).

At the same time, comparison of the electron phase diagrams for LaFeAsO$_{1-x}$F$_x$ and other ⟨1111⟩ compounds (SmFeAsO$_{1-x}$F$_x$ and CeFeAsO$_{1-x}$F$_x$) [30] demonstrates significant discrepancies in the positions of the superconducting domes. These discrepancies, as shown in [31], appear to be due to a difference in the real content of fluorine in specimens of SmFeAsO$_{1-x}$F$_x$ and CeFeAsO$_{1-x}$F$_x$ from the nominal content determined by the initial weight. At the same time, in LaFeAsO$_{1-x}$F$_x$ the real content of fluorine coincides with the nominal value. With respective corrections made, the regions of the superconducting domes on the phase diagrams of ⟨1111⟩ compounds coincide [31].

## 2.6. $Ba_{1-x}K_xFe_2As_2$

Consider now the hole-doped compound $Ba_{1-x}K_xFe_2As_2$. Substitution of K for Ba leads to the emergence in one of the two FeAs planes of a hole that is distributed over four nearest As ions (Fig. 11a), imparting to them charge $q^*$. (Localization of the doped holes on As ions is indirectly confirmed by the results of [32], where the contribution of electron carriers to thermal conductivity was observed in $Ba_{1-x}K_xFe_2As_2$ up to $x = 0.88$.) Thus, four CT plaquettes are formed around each projection of K ion.

The percolation cluster of CT plaquettes can be produced at an ordered arrangement of dopants into a $\sqrt{10} \times \sqrt{10}$ lattice (Fig. 11b). The concentration conforming to the total filling of the $\sqrt{10} \times \sqrt{10}$ lattice is $x = 0.2$ (with account for the fact that only each second K ion dopes a hole to this FeAs plane), and the concentration conforming to the percolation threshold, $x = 0.12$. The most optimal is the ordering of dopant projections into the $\sqrt{5} \times \sqrt{5}$ lattice (Fig. 11c), which corresponds to an optimal dopant concentration $x = 0.4$. The concentration values determined in this way are well consistent with the phase diagram of $Ba_{1-x}K_xFe_2As_2$ (Fig. 2f [8]).

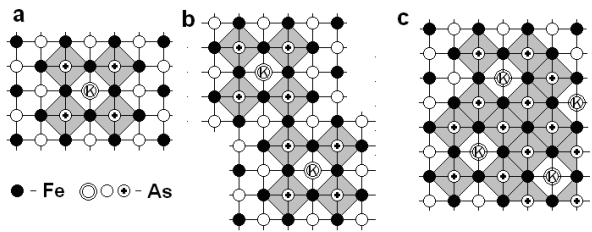

Figure 11. (a) Formation of CT plaquettes in $Ba_{1-x}K_xFe_2As_2$. Open circles indicate projections of As ions at the vertices of the tetrahedra onto the Fe plane; a double circled K symbol is the projection of an As ion coinciding with the projection of a K ion on the Fe plane and being the centre of CT plaquette (shaded); circled plus symbols are As ions carrying charge $q^*$; (b) formation of a percolation cluster of CT plaquettes on the $\sqrt{5} \times \sqrt{5}$ square lattice.

## 2.7. $Ca_{1-x}La_xFe_2As_2$

In this compound, substitution of La for Ca leads to the emergence of an additional electron in one of the two FeAs planes that imparts charge $q^*$ ($\approx -|e|/4$) to each of four Fe ions nearest to the projection of La (Fig. 12). As the result, eight CT plaquettes form around these Fe ions.

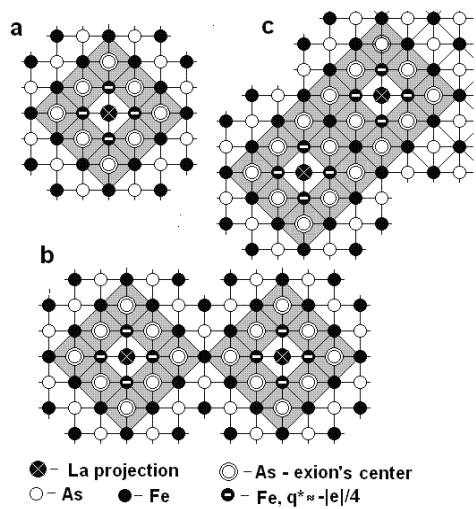

Figure 12. (a) Formation of CT plaquettes in $Ca_{1-x}La_xFe_2As_2$. Eight CT plaquettes (shaded) form around a La projection in one of the FeAs planes; open circles, As; filled circles, Fe; filled minus circles, Fe ions with charge $q^* = -|e|/4$. (b) Formation of a percolation cluster of CT plaquettes on the $\sqrt{18} \times \sqrt{18}$ superlattice; (c) formation of a percolation cluster of CT plaquettes on the 3×3 superlattice.

The maximal distance between dopant projections, required to form a percolation cluster of CT plaquettes, is $l_D = \sqrt{18}$ (Fig. 12b). The percolation threshold at the ordering of dopant projections to a $\sqrt{18} \times \sqrt{18}$ superlattice corresponds to a concentration of $x = 0.067$ (with account for the fact that only each second La ion dopes an electron to the FeAs plane).

Optimal doping, as seen in Fig. 12c, is in correspondence with the complete ordering of dopant projections to the 3×3 superlattice. A concentration corresponding to an optimal doping is $x = 0.22$; respectively, a concentration that conforms to the percolation threshold at this lattice is $x = 0.132$. All three concentration values (0.067, 0.13 and 0.22) coincide with the boundaries for the low- and high-temperature superconductivity regions on the experimental phase diagram of $Ca_{1-x}La_xFe_2As_2$ (Fig. 2g [9]).

Thus, we have shown that the regions of concentrations corresponding to the superconducting dome on the phase diagrams of cuprates and pnictides coincide with those of the formation of a percolation cluster of the CT phase[2]. The demonstrated coincidence can serve as an indication of an improper approach that considers cuprates and pnictides as spatially homogeneous systems with concentrations of carriers determined by the level of doping. The fact that the proposed method of constructing phase diagrams proved similarly successful both for cuprates and pnictides is a serious argument in favour of the common nature of the superconducting state in these classes of HTSCs.

## 3. Heitler–London centres and generation of free carriers

Let us return to consider the properties of the phase with $\Delta_{ct} = 0$ where a CT plaquette is formed on each cation (in cuprates) or anion (in pnictides). Let there be two CT plaquettes centred at the nearest Cu cations (in cuprates) or As anions (in pnictides). Such a pair of plaquettes forms a Heitler–London (HL) centre, which can be considered as a solid state analog of the hydrogen molecule [21]. At this centre, two electrons and two holes can form a bound state (biexciton) that is energetically lower than the energy of two CT excitons due to the possibility of two holes (electrons) in singlet state to be in space between the central ions and to be attracted simultaneously to two electrons (holes) occurring on these ions. An additional decrease of energy, $\Delta E_{HL}$, can in this case be assessed from the ratio $\Delta E_{HL} \sim \Delta E_{H_2}/\varepsilon_\infty^2 \approx 0.2$ eV, where $\Delta E_{H_2} = 4.75$ eV is the binding energy in the $H_2$ molecule and $\varepsilon_\infty \approx 4.5–5$ (for cuprates) [33]. An HL centre will be considered to be occupied if both electron (hole) states on the central cations of Cu (anions of As) are occupied.

Thus, in a CT phase two electrons on Cu cations (two holes on As anions) will have a lower energy if they occupy adjacent cations (anions) and two holes (two electrons) bound to them are on the surrounding O anions (Fe cations). These pairs of electrons (holes) occupy a common paired level of HL centres that belong to one percolation cluster (Fig. 13). In this figure, the states of electrons in band $L_d$ (in cuprates) or holes in band $d_L$ (in pnictides) represent a superposition of band and exciton states.

At an increase of temperature, the biexciton can become ionized without being separated into two excitons. By analogy with the hydrogen molecule, apart from the $H_2$ molecule, there is also the bound state of two protons and one electron, namely the $H_2^+$ ion. In our case, this corresponds to an ionized HL centre, at which two electrons and one hole (in cuprates) or two holes and one electron (in pnictides) form a bound state. The emerging "spare" hole (in cuprates) or electron (in pnictides) can pass from node to node, thus overlapping the corresponding orbitals of HL centres and forming a hole (electron) subband with the corresponding type of superconductivity on the level of chemical potential $\mu$ (Fig. 13).

---

[2] Note that the consideration of the case of HTSCs with heterovalent doping does not rule out the possible existence of compounds where the percolation exion cluster already exists in the undoped phase (e.g., LiFeAs), or forms during isovalent substitution, if the dopant locally decreases $\Delta_{ib}$ and becomes an exion centre (e.g., $BaFe_2(As_{1-x}P_x)_2$).

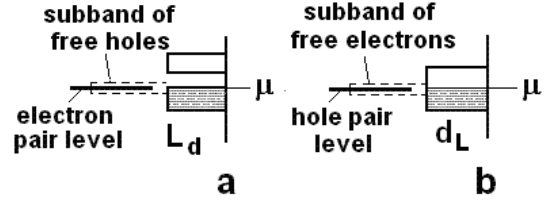

Figure 13. One-electron band structure schemes of (a) cuprates (in electron representation) and (b) pnictides (in hole representation). μ, the level of chemical potential. The electron (hole) states in the $L_d$ band ($d_L$ band) represent a superposition of the band and exciton states. The electron (hole) band of HL centres is formed at $T > 0$ as the result of their filling with electrons (holes).

The concentration of free carriers in the emerging subband of HL centres (holes, in cuprates; electrons, in pnictides) will be determined by HL-centre occupation, i.e., by the balance between the rates of transition of electron (hole) pairs to HL centres and the rates of their departure to the band as the result of the breakdown of a pair state. The value of pair hybridization Γ (or the inverse lifetime of a pair state) depends on temperature [34,35]:

$$\Gamma \approx kT \cdot (V/E_F)^2,$$

(here $V$ is the single-particle hybridization constant; $E_F$, Fermi energy; $T$, temperature). The rate of pair level–band transitions is $\propto \eta\Gamma$. The inverse process (occupation of a pair level) is determined by the rate of electron–electron scattering; its rate is $\propto (1-\eta)T^2$. Hence, for the occupation of an HL centre, η, we have:

$$\eta = 2T/(T+T_0), \qquad (1)$$

where $T_0$ is a temperature-independent constant. Correspondingly, for the concentration of additional carriers $n = 2NT/(T + T_0)$, where $N$ is the concentration of HL centres. From this relation, it follows that at $T = 0$ the concentration of additional free carriers $n = 0$. (Here we assume that at $T = 0$ the occupation of HL centres is negligibly small.) The maximally possible concentration of additional carriers is achieved at the total filling of HL centres; i.e., if an CT plaquette cluster fills the entire basal plane, this concentration will equal the concentration of Cu (As) ions in the basal planes. That is, for $YBa_2Cu_3O_7$, which have no free carriers at $T = 0$, $n \to 2$ (u-cell)$^{-1}$ at $T \to \infty$, which corresponds to the experiment [10,36].

According to the above consideration, noncoherent transport in the normal state occurs due to carriers that emerge upon filling HL centers. However, at T = 0 the pair level is not filled, and the concentration of carriers is n=0. Consequently, at T = 0 noncoherent transfer would be impossible in view of the absence of free carriers. At the same time, in such a system, where there are an electron and a hole in each CT plaquette (i.e., there are free states to which electrons can pass), coherent transport becomes possible when all carriers of a like sign move in a given direction coherently, as a whole (e.g., superconducting condensate). The latter is possible in the presence of pairing interaction.

As shown in [34,35,37–41], the account for scattering processes with intermediate bound states lying in the vicinity of $E_F$ can lead to a strong renormalization of efficient electron–electron interaction capable of providing for high $T_c$ in the system. In a pattern based on the formation of local CT excitons and HL centres, electron–electron attraction emerges due to the formation of a bound state of two electons (holes) getting at the central cations (anions) of an unfilled HL centre and two holes (electrons), with necessity emerging on the surrounding ions in the phase with $\Delta_{ct} = 0$. Thus, the basal planes of doped cuprates and pnictides can be considered as one more type of structures (in addition to one-dimensional Little chains [42] and Ginzburg sandwiches [43]), where the exciton mechanism of superconductivity can be realized.

## 5. Conclusion

This article reviews experimental phase diagrams of cuprates and pnictides to demonstrate that specific features of the superconducting phase diagrams in both HTSC families can be understood within the framework of the approach that assumes the localization of doped carriers in the vicinity of dopants as a consequence of their forming the localized trion complexes consisting of a doped carrier and its surrounding charge transfer (CT) excitons. This leads to the formation of hydrogen-like ion complexes, CT plaquettes, in the vicinity of the dopant; a CT exciton resonantly interacting with band states can form in these CT plaquettes.The geometry of CT plaquettes' arrangement in the basal plane of the crystal is determined by its crystalline structure and type of dopant; therefore, the dopant concentration range corresponding to the existence of a percolation cluster of CT plaquettes can be readily determined for each particular compound. For a number of well investigated HTSCs, the indicated concentration ranges are well consistent with the experimental values.

A close proximity of doping impurities leads to the emergence of paired plaquette complexes (HL centres) that are in the basal plane and represent pairs of adjacent ions of one sign surrounded by ions of the other sign (a solid state analog of the hydrogen molecule). At this centre, two electrons and two holes can form a bound biexciton state owing to the possibility of two holes (in cuprates) or two electrons (in pnictides) in singlet state to be in space between the central ions and be attracted simultaneously to two electrons (holes) on these ions. The existence of such HL centres makes it possible to explain the generation of additional free carriers in the normal state and the emergence of superconducting pairing.

**Acknowledgements**
The authors are grateful to A.A. Gorbatsevich for useful discussions and critical comments. The work was supported by the RFBR (grant No 14-02-0078516).